\documentclass[aps,prl,twocolumn,superscriptaddress]{revtex4-2}

\usepackage{graphicx}                   % Include figure files
\usepackage{dcolumn}                    % Align table columns on decimal point
\usepackage{mathtools}
\usepackage[colorlinks=true]{hyperref}  % add hypertext capabilities
\usepackage[dvipsnames]{xcolor}         % citation colors

\setcitestyle{super}

\newcommand\myshade{85}
\colorlet{mylinkcolor}{Red}
\colorlet{mycitecolor}{Blue}
\colorlet{myurlcolor}{Gray}

\hypersetup{
  linkcolor  = mylinkcolor!\myshade!black,
  citecolor  = mycitecolor!\myshade!black,
  urlcolor   = myurlcolor!\myshade!black,
  colorlinks = true,
}

\newlength{\thelinewidth}
\thelinewidth=\textwidth

\begin{document}
\title{Evidence of dual Shapiro steps in a Josephson junctions array}

\author{Nicolò Crescini}
    \email{nicolo.crescini@neel.cnrs.fr}
    \thanks{the authors contributed equally to this work.}
    \affiliation{Univ. Grenoble Alpes, CNRS, Grenoble INP, Institut N\'eel, 38000 Grenoble, France}
\author{Samuel Cailleaux}
    \email{samuel.cailleaux@neel.cnrs.fr}
    \thanks{the authors contributed equally to this work.}
    \affiliation{Univ. Grenoble Alpes, CNRS, Grenoble INP, Institut N\'eel, 38000 Grenoble, France}
\author{Wiebke Guichard}
	\affiliation{Univ. Grenoble Alpes, CNRS, Grenoble INP, Institut N\'eel, 38000 Grenoble, France}
\author{Cécile Naud}
	\affiliation{Univ. Grenoble Alpes, CNRS, Grenoble INP, Institut N\'eel, 38000 Grenoble, France}
\author{Olivier Buisson}
	\affiliation{Univ. Grenoble Alpes, CNRS, Grenoble INP, Institut N\'eel, 38000 Grenoble, France}
\author{Kater Murch}
    \affiliation{Department of Physics, Washington University, St. Louis, Missouri 63130, USA}
\author{Nicolas Roch}
    \email{nicolas.roch@neel.cnrs.fr}
	\affiliation{Univ. Grenoble Alpes, CNRS, Grenoble INP, Institut N\'eel, 38000 Grenoble, France}
	
\date{\today}

% introductory paragraph
\begin{abstract}
The modern primary voltage standard is based on the AC Josephson effect and the ensuing Shapiro steps, where a microwave tone applied to a Josephson junction yields a constant voltage $hf/2e$ ($h$ is Planck's constant and $e$ the electron charge) determined by only the microwave frequency $f$ and fundamental constants \cite{josephson_possible_1962,shapiro_josephson_1963}.
Duality arguments for current and voltage  \cite{schon_quantum_1990,spiller_electromagnetic_1990,ingold_charge_1992,corlevi_phase-charge_2006,arutyunov_superconductivity_2008,guichard_phase-charge_2010,kerman_fluxcharge_2013} have long suggested the possibility of dual Shapiro steps---that a Josephson junction device could produce current steps with heights determined only on the applied frequency \cite{likharev_theory_1985,averin_bloch_1985,AVERIN1990935,hu_bloch_1993}. In this report, we embed an ultrasmall Josephson junction in a high impedance array of larger junctions to reveal dual Shapiro steps \cite{guichard_phase-charge_2010}.
For multiple frequencies, we detect that the AC response of the circuit is synchronised with the microwave tone at frequency $f$, and the corresponding emergence of flat steps in the DC response with current $2ef$, equal to the tunnelling of a Cooper pair per tone period.
This work sheds new light on phase-charge duality, omnipresent in condensed matter physics\cite{Shahar.1996,Ovadia.2013}, and extends it to Josephson circuits. Looking forward, it opens a broad range of possibilities for new experiments in the field of circuit quantum electrodynamics \cite{puertas_martinez_tunable_2019, pechenezhskiy_superconducting_2020} and is an important step towards the long-sought closure of the quantum metrology electrical triangle \cite{likharev_bloch_1985,pekola_single-electron_2013,bylander_current_2005}.%Looking forward, it opens a broad range of possibilities for new experiments, e.\,g. in the field of circuit quantum electrodynamics \cite{puertas_martinez_tunable_2019, pechenezhskiy_superconducting_2020}. It is also an important step towards the long-sought closure of the quantum metrology electrical triangle \cite{likharev_bloch_1985,pekola_single-electron_2013,bylander_current_2005}.
\end{abstract}

\maketitle

    \begin{figure*}
    \centering
    \includegraphics[width=.925\textwidth]{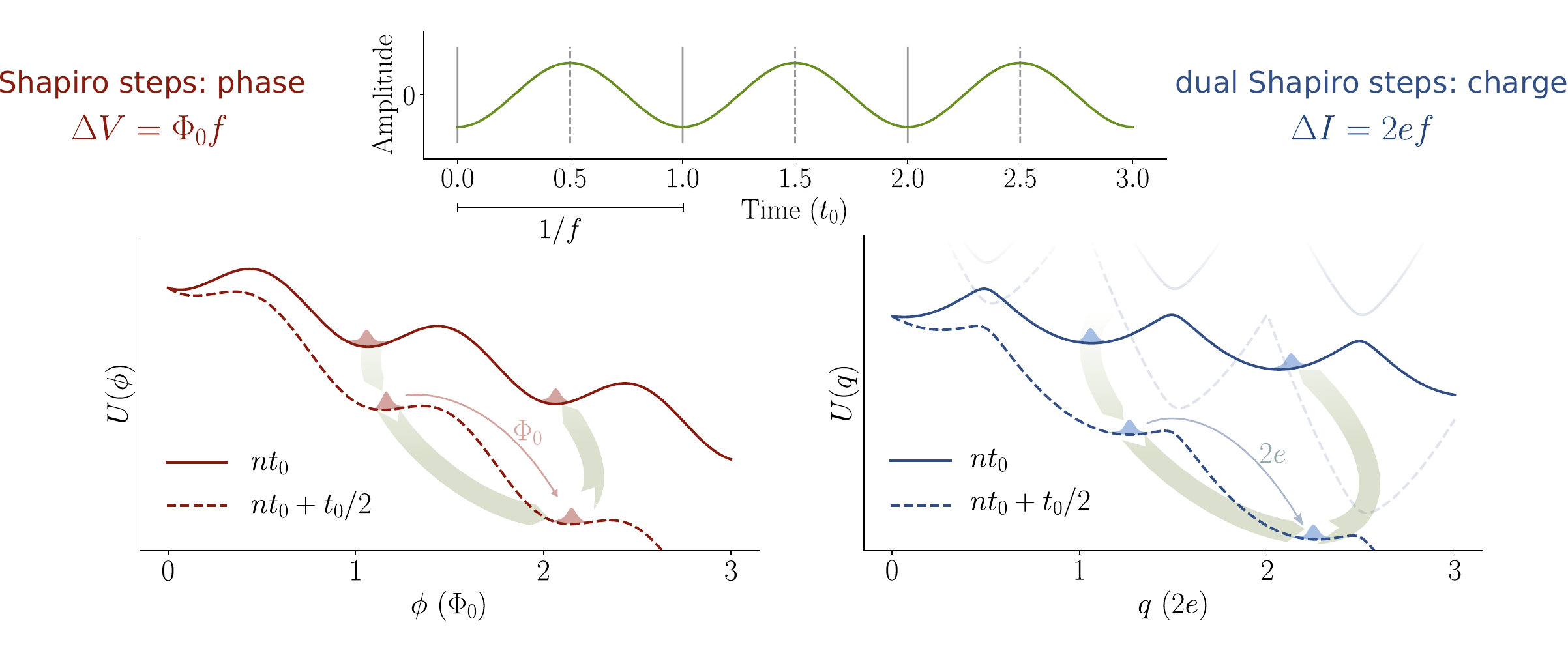}
    \caption{%The quantum metrology electrical triangle. Defined by caesium atomic clocks, a frequency, $f$, measured in Hertz is the timekeeper of the International System of Units. The frequency of a microwave tone can be synchronised with two coherent quantum effects to define the units of voltage and current at the base of the triangle.
    The Shapiro steps and their dual. Usual Shapiro steps are realised by periodically tilting the washboard potential of a Josephson junction with $E_J\gg E_C$.
    %The voltage standard (left side) is realised by periodically tilting the washboard potential of a Josephson junction with $E_J\gg E_C$.
    %For a $E_J\gg E_C$ (left side), the phase $\phi$ across the Josephson junction is a well defined parameter and the energetics of the circuit are given by a tilted washboard potential. 
    Since the phase is a well defined variable, the driving of the potential yields steps of constant voltage $\Phi_0 f$ ($\Phi_0=h/2e$ is the flux quantum) that are precisely determined by the metered transport of fluxons.
    %Since the phase is a well defined variable, the driving of the potential yields Shapiro steps of constant voltage $\Phi_0 f$ ($\Phi_0=h/2e$ is the flux quantum) that are precisely determined by the metered transport of fluxons.
    %The Josephson effect yields Shapiro steps of constant voltage that are determined precisely by an applied frequency, and are used to define the Volt. 
    In contrast, the dual phenomenon (right side) uses the lowest Bloch band of a junction with $E_J\le E_C$ (higher bands in the picture are shown only for an illustrative purpose).
    %In contrast, its dual phenomenon (right) uses a junction with $E_J\le E_C$ that yields an energetic landscape of Bloch bands, higher bands in the picture are shown only for an illustrative purpose. 
    With charge as the well defined variable, current steps of value $2ef$ determined by the cyclic transport of Cooper pairs appear: the dual Shapiro steps.
    %With charge as the well defined variable, dual Shapiro steps---current steps of value $2ef$ determined by the cyclic transport of Cooper pairs---can give the metrological definition of the Ampère. %Ultimately the connection between the Volt and the Ampère is established by the quantum Hall effect (bottom).
    }
    \label{fig:triangle}
    \end{figure*}

In quantum mechanics, uncertainty relations quantify the incompatibility of quantum operators. For a Josephson junction these operators are the charge $Q$ and the Josephson phase $\phi$, which respect $[\phi, Q]=2ie$, where $e$ is the electron charge \cite{tinkham2004introduction}.
Because of this incompatibility, the suppression of the variance of one of these operators causes the conjugate operator's variance to increase.
The balance between phase and charge fluctuations in a junction is determined by the ratio of two energy scales, the Josephson energy $E_J=\hbar I_c/2e$ and the charging energy $E_C=e^2/2C$, where $I_c$ is the critical current of the junction, $C$ is its capacitance and $\hbar$ is the reduced Plank constant\cite{tinkham2004introduction}.
In one limit, $E_J/E_C\gg1$, the variance of $\phi$ is small, allowing it to be treated as a classical variable, and the dynamics of the system is described by a tilted washboard potential $U(\phi)$ shown in Fig.\,\ref{fig:triangle}. By applying a microwave tone, the potential is periodically tilted; one fluxon ($\Phi_0$) is transported across the junction at each period resulting in discrete voltage steps---Shapiro steps---in the IV curve.
Au contraire, if $E_J/E_C\lesssim 1$, fluctuations in the phase are enhanced allowing the charge variance to be reduced. 
In this limit, one expects a microwave drive to produce \emph{dual} Shapiro steps; the cyclical tilting of a potential periodic in charge can meter the transport of Cooper pairs across the junction, resulting in current steps proportional to the applied frequency. 
	\begin{table}[b]
	Dual transformations\\
	\vspace{0.1cm}
	\begin{tabular}{ l | l  }
	%Thing 	            &	Dual thing              \\
	\hline
	$E_J/E_C\gg1$                   &   $E_J/E_C\le1$               \\
	Classical (localised) $\phi$    &   Classical (localised) $q$   \\
	Tilted washboard $U(\phi)$      &   First Bloch band $U(Q)$     \\
	Nonlinear inductance            &   Nonlinear capacitance       \\
	Critical current $I_c$          &   Critical voltage $V_c$      \\
	AC Josephson effect             &	Bloch oscillations	        \\
	Fluxons transport               &   Cooper pairs transport      \\
	Shapiro steps                   &   Dual Shapiro steps          \\
	\end{tabular}
	\caption{Summary of dual transformations in a Josephson circuit.}
	\label{tab:duality}
    \end{table}

An ultrasmall Josephson junctions (UJJ) has an $E_J/E_C$ ratio smaller than one.
The junction Hamiltonian is
\begin{equation}
H = 4E_C \left(Q/2e\right)^2 - E_J\cos(\phi),
\label{eq:hamiltonian}
\end{equation}
which, using Bloch's theorem\cite{Ashcroft:102652}, can be recast to
\begin{equation}
    H=\sum_s U^{(s)}(q),
\end{equation}
where we have introduced the quasicharge $q$ and the Bloch bands $U^{(s)}(q)$\cite{likharev_theory_1985,likharev_theory_1985}.
When this junction is embedded in an environment of impedance $R$, thermal and quantum fluctuations can further increase the charge variance. Here, charge fluctuations are bounded by the energy-time uncertainty relation $\Delta E \Delta \tau >  \hbar$, where the charging time of the junction capacitance is $\Delta \tau= RC$. Hence, large impedance is necessary to reduce the charge fluctuations, in particular to have $\Delta E<E_C$ requires, $R>h/4 e^2\simeq6.45\,\mathrm{k\Omega}$, the quantum of resistance. Additionally, to suppress thermal fluctuations one has to cool the environment to ultracryogenic temperatures $T<E_C/k_\mathrm{B}$.
Under these conditions the charge is localised and thus can be treated as a classical variable. $Q$ is well approximated by $q$ as its dynamics is restricted to the first Bloch band ($s=0$), thereby realising the periodic potential $U(Q)\simeq U^{(0)}(q) = \cos(\pi Q / e)$---dual to $U(\phi)$---shown in Fig.\,\ref{fig:triangle}. %To first order, the potential is approximated as $U(Q) \simeq E_Q \cos(\pi Q / e)$, 
The periodic potential corresponds to a nonlinear capacitance---dual to the nonlinear inductance typically associated with a Josephson junction. Table \ref{tab:duality} summarises the dual transformations in Josephson circuits.
Along these lines, the amplitude of the first Bloch band is related to the critical voltage $V_c = \pi E_Q/e$, the minimum voltage required for a flow of DC current across the junction.
$V_c$ is dual to the critical current, the current below which no voltage is generated\cite{corlevi_phase-charge_2006}. Upon the application of a DC current, the system exhibits Bloch oscillations\cite{likharev_theory_1985,kuzmin_observation_1991}, i.e. voltage oscillations arising in the presence of a constant force, a phenomenon dual to the AC Josephson effect.
If these oscillations are synchronised with an external tone, they yield dual Shapiro steps\cite{likharev_theory_1985,guichard_phase-charge_2010,di_marco_quantum_2015}. 

Prior work has demonstrated some of the essential components for the exploration of Josephson circuits' duality. A charge periodic potential has been obtained in UJJs\cite{kuzmin_observation_1991,andersson_synchronous_2000,weisl_bloch_2015,cedergren_insulating_2017} or nanowires \cite{lau_quantum_2001,lehtinen_coulomb_2012,doi:10.1063/1.5092271,PhysRevB.87.144510}. Large impedance environments have been realised by using resistors \cite{kuzmin_observation_1991}, suspended \cite{peruzzo_geometric_2021} or low dielectric constant \cite{Rolland.2019} substrates, high kinetic inductance materials \cite{yoshihiro_observation_1988,lau_quantum_2001,PhysRevB.87.144510,grunhaupt_granular_2019}, or metamaterials \cite{Bell.2012, Masluk.2012, puertas_martinez_tunable_2019,leger_observation_2019,pechenezhskiy_superconducting_2020}. Despite this progress a clear demonstration of dual Shapiro steps is still missing. In this work, we harness techniques from circuit QED\cite{RevModPhys.93.025005} to study the interplay of microwave drive and DC bias of an UJJ embedded in a superconducting high-impedance environment. Utilising low noise microwave spectroscopy, we observe the onset of a microwave mode synchronised with an external tone. In turn, we observe current plateaux proportional to the frequency of the microwaves. 
These two experimental observations are consistent with synchronised Bloch oscillations and dual Shapiro steps.

	\begin{figure*}
	\centering
	\includegraphics[width=.8\textwidth]{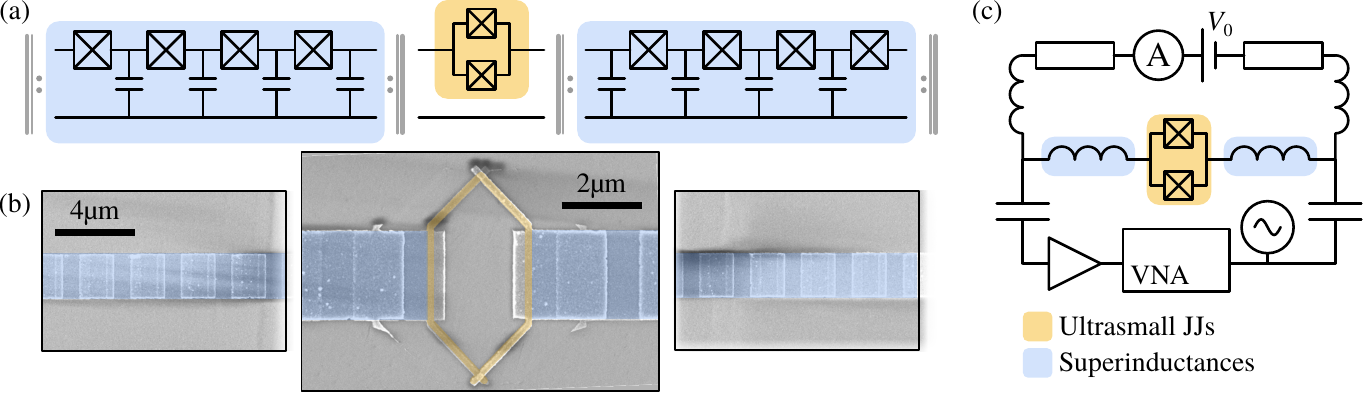}
	\caption{Overview of device and measurement setup, where the UJJ is shown in orange and the JJ array in blue. (a) Circuit schematic of the Bloch array. (b) SEM image of  the UJJ and of the beginning of JJ arrays nominally identical to the one under test. (c) Simplified schematic of the measurement setup including the device, bias tees, damping resistors, DC and RF electronics, and a vector-network analyser (VNA).}
	\label{fig:circs}
	\end{figure*}
	
Fig.\,\ref{fig:circs} shows the device used in this work, which we refer to as a Bloch array, and consists in a superconducting quantum interference device (SQUID) formed by two ultrasmall junctions surrounded by two linear JJs superinductances which provide the high impedance environment. 
The array consists of Al/AlO$_x$/Al junctions in stripline geometry on a fused silica wafer, which reduces the capacitance to ground \cite{arndt_dual_2018}.
The SQUID is formed by junctions of area $0.2\times0.2\,\mathrm{\mu m^2}$ ($E_J/E_C\simeq0.4$) each, and the superinductance comprises an array of $N_\mathrm{a}/2=1750$ junctions with an area of $1\,\mathrm{\mu m^2}$ ($E_J/E_C\simeq250$) \cite{largesmall}.
From these parameters we estimate a first Bloch band width of $E_Q/h=2.5\,\mathrm{GHz}$ and a minimum gap of $7.9\,\mathrm{GHz}$ between the first and the second Bloch band. Each superinductance's characteristic impedance and inductance are $Z=8.0\,\mathrm{k\Omega}$ and $L/2=3.3\,\mathrm{\mu H}$, respectively.
A scanning electron microscope (SEM) image is shown in Fig.\,\ref{fig:circs}b. Further details are given in the Methods section. 
The Bloch array is studied with an experimental setup allowing for simultaneous DC and RF measurements, as is shown in Fig.\,\ref{fig:circs}c. Measurements are performed at the base temperature of a dilution refrigerator with $T\simeq 23\,\mathrm{mK}$.

	\begin{figure*}
	\centering
	\includegraphics[width=.98\textwidth]{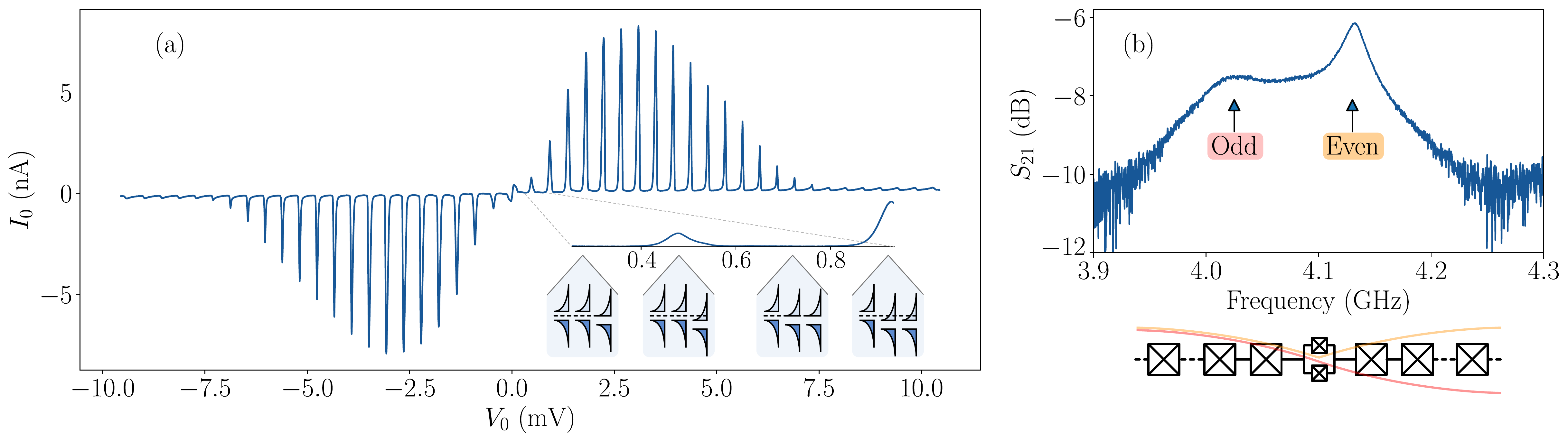}
	\caption{DC and RF characterisation of the Bloch array. Plot (a) is the low voltage IV curve, which is detailed in the text. The inset schematically shows the switching mechanism of the junctions in the array with a band model, where the dashed line is a guide to the eye. The microwave transmission of the Bloch array's fourth mode doublet is displayed in (b), where the voltage profile of odd and even modes is shown below.}
	\label{fig:dcrf}
	\end{figure*}
	
We first examine the DC and RF response of the Bloch array separately.
The IV curve is obtained by biasing the device with the voltage $V_0$ and reading out the corresponding current $I_0$. 
Here we consider $|V_0|\le10\,\mathrm{mV}$ while the whole IV curve is described in the Methods section. As is shown in the plot, the current-voltage characteristics exhibits several current peaks\cite{haviland_superconducting_2000,dolata_single-charge_2005} evenly spaced by $2\Delta$, the superconducting gap of aluminium. These peaks can be understood as follows. When a voltage is applied to the Bloch array one of the junctions ($\mathrm{J_1}$) is biased to the sub-gap state, behaving as a large series resistance, and the flow of current is suppressed.
Consequently, $V_0$ drops on $\mathrm{J_1}$ until it reaches the gap.
When $V_0=2\Delta$, $\mathrm{J_1}$ turns normal conducting, and $I_0$ sharply increases until a second junction $\mathrm{J_2}$ is biased into the sub-gap state, introducing again a large series resistance and suppressing the current.
To turn both the junctions resistive the voltage needs to be $4\Delta$, and the process repeats. A schematic of this mechanism is reported in the inset of Fig.\,\ref{fig:dcrf}a.
This successive switching resets the voltage drop across the junctions every $2\Delta$. In this way the same physics, and in particular the low voltage physics, repeats at voltages modulo $2\Delta$ as one more junction switches to the normal state per step.
% including one more junction per step, giving origin to a replica symmetry \comment{NC-need to verify the terminology}. 
RCSJ simulations of a voltage biased JJ array, reported in the Methods, are in agreement with our observations.

\begin{figure*}
	\centering
	\includegraphics[width=\textwidth]{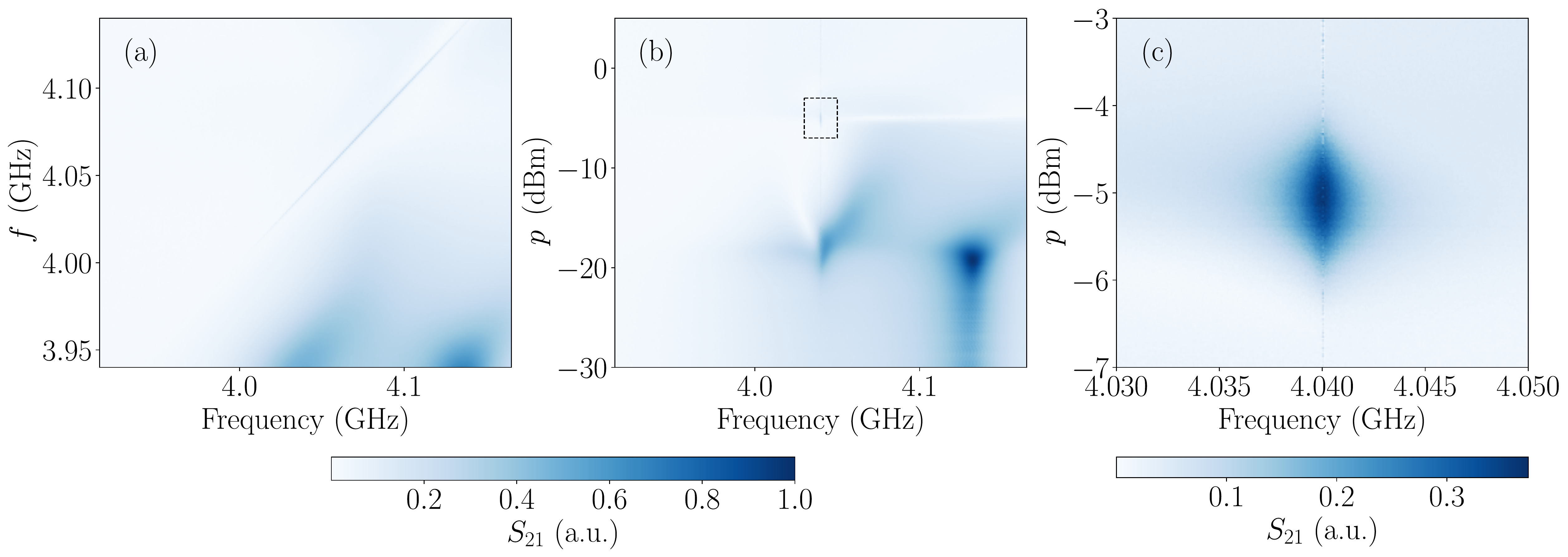}
	\includegraphics[width=.8\textwidth]{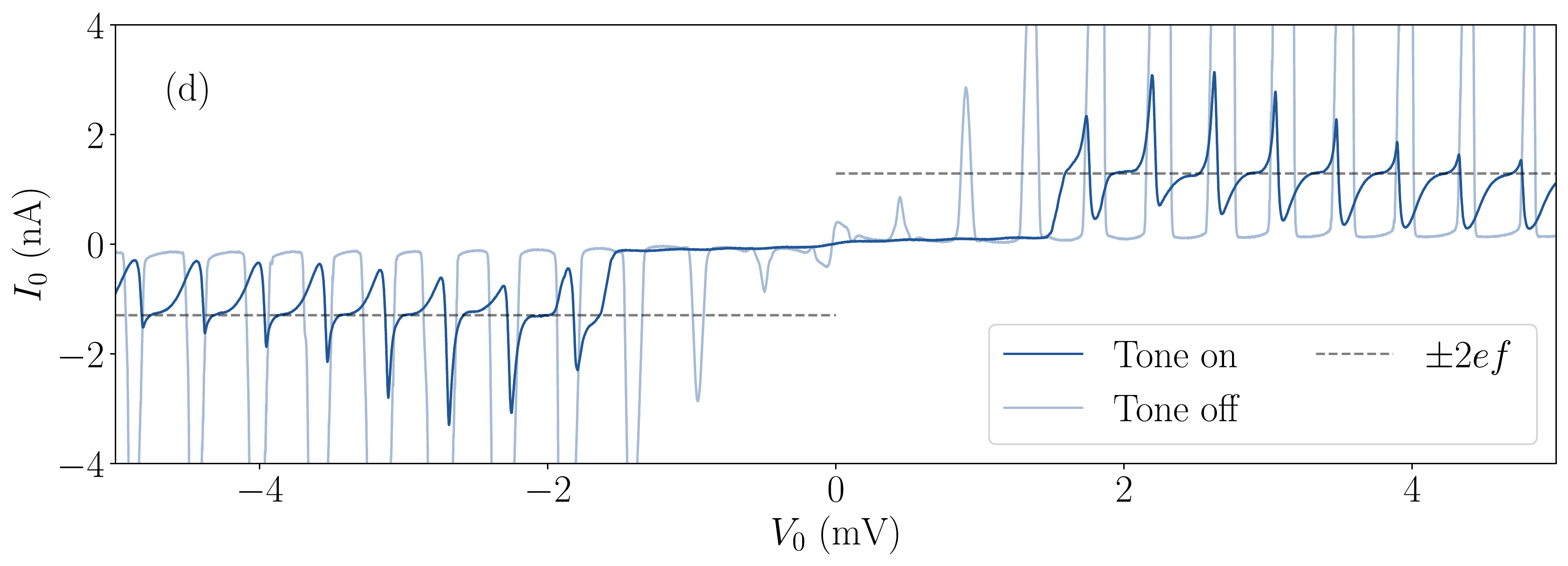}
	\caption{Emergence of Bloch oscillations and dual Shapiro steps. In (a) the frequency of the external tone $f$ is changed to observe the effect it has on the transmission spectrum of the Bloch array. A similar measurement is reported in (b) where we vary the power $p$ of the tone at fixed frequency $f=4.04\,\mathrm{GHz}$. The dashed rectangle evidences the plot region reported in (c), where the transition between tone interference and microwave mode is clearly visible. Plot (d) shows the IV characteristics measured for $f=4.04\,\mathrm{GHz}$ and $p=-5.0\,\mathrm{dBm}$. Between two current peaks the plateaux current is $\pm2ef$, as is evidenced by the grey dashed lines. 
	%The data did not undergo any post-processing.
	}
	\label{fig:pfsweep}
\end{figure*}

The microwave transmission of a Bloch array exhibits mode doublets, one with odd and one with even symmetry \cite{leger_observation_2019}, as is shown in Fig.\,\ref{fig:dcrf}b. 
The odd mode is coupled to the UJJ at the centre of the array, while the even one is not, as is comprehensively discussed by Leger et al.\cite{leger_observation_2019}.
The frequency of the odd mode is lower than the even mode, which is consistent with the presence of a capacitive element between the left and right superinductances. This agrees with duality considerations, according to which the UJJ should behave as a nonlinear capacitance.
Leveraging the DC and microwave capabilities of our setup, we investigate the effect of an applied voltage on the modes of the Bloch array. By measuring the $S_{21}$ as a function of $V_0$, we observe that between two current peaks the doublet is visible yet starts to degrade when current flows through the array.
The mode doublet is replicated after every $2\Delta$-switch, compatible with the mechanism depicted by the IV curve. The full measurement is detailed in the Methods and in Fig\,\ref{fig:s21_dc}.

	\begin{figure*}
	\centering
	\includegraphics[width=.39\textwidth]{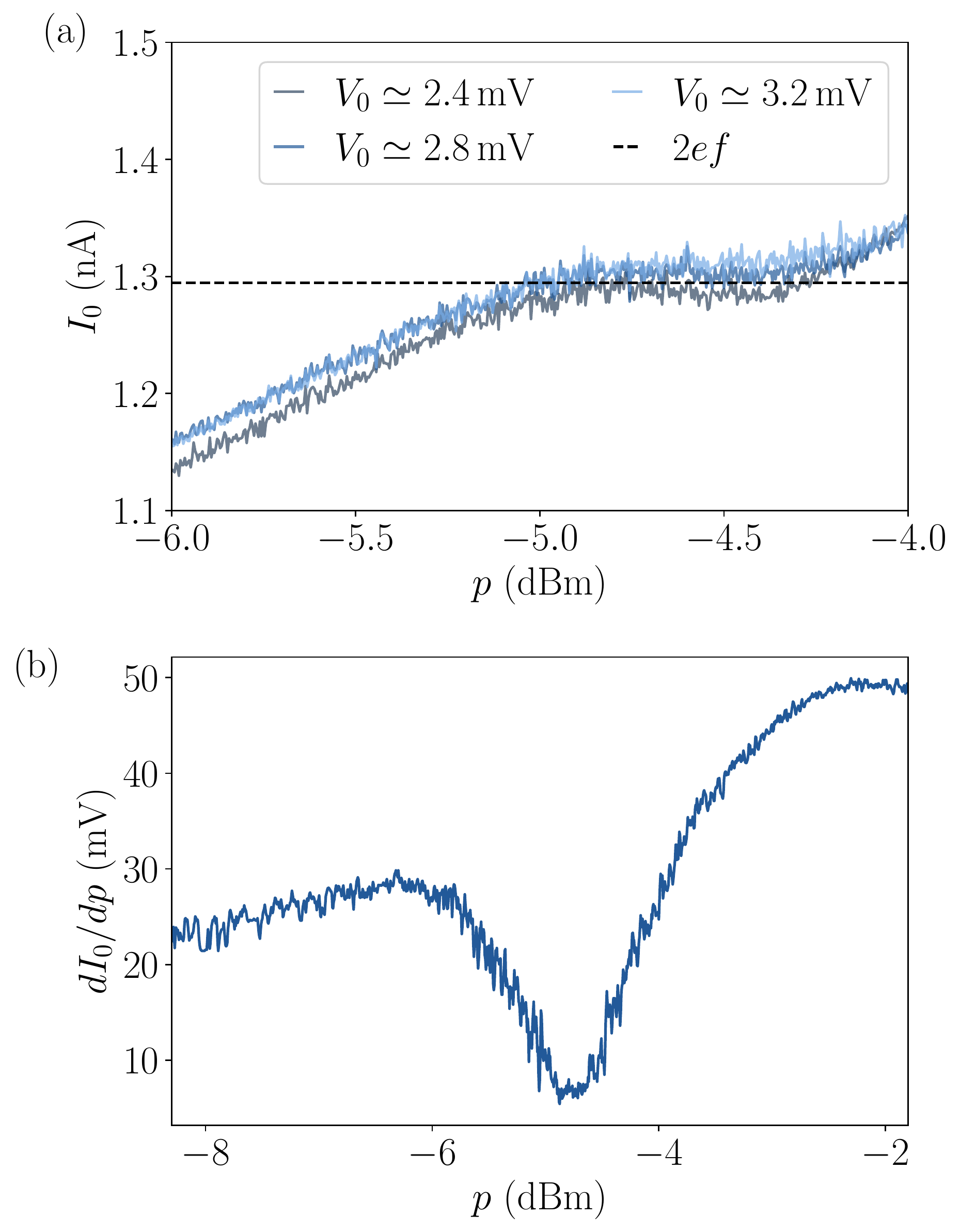}
	\includegraphics[width=.59\textwidth]{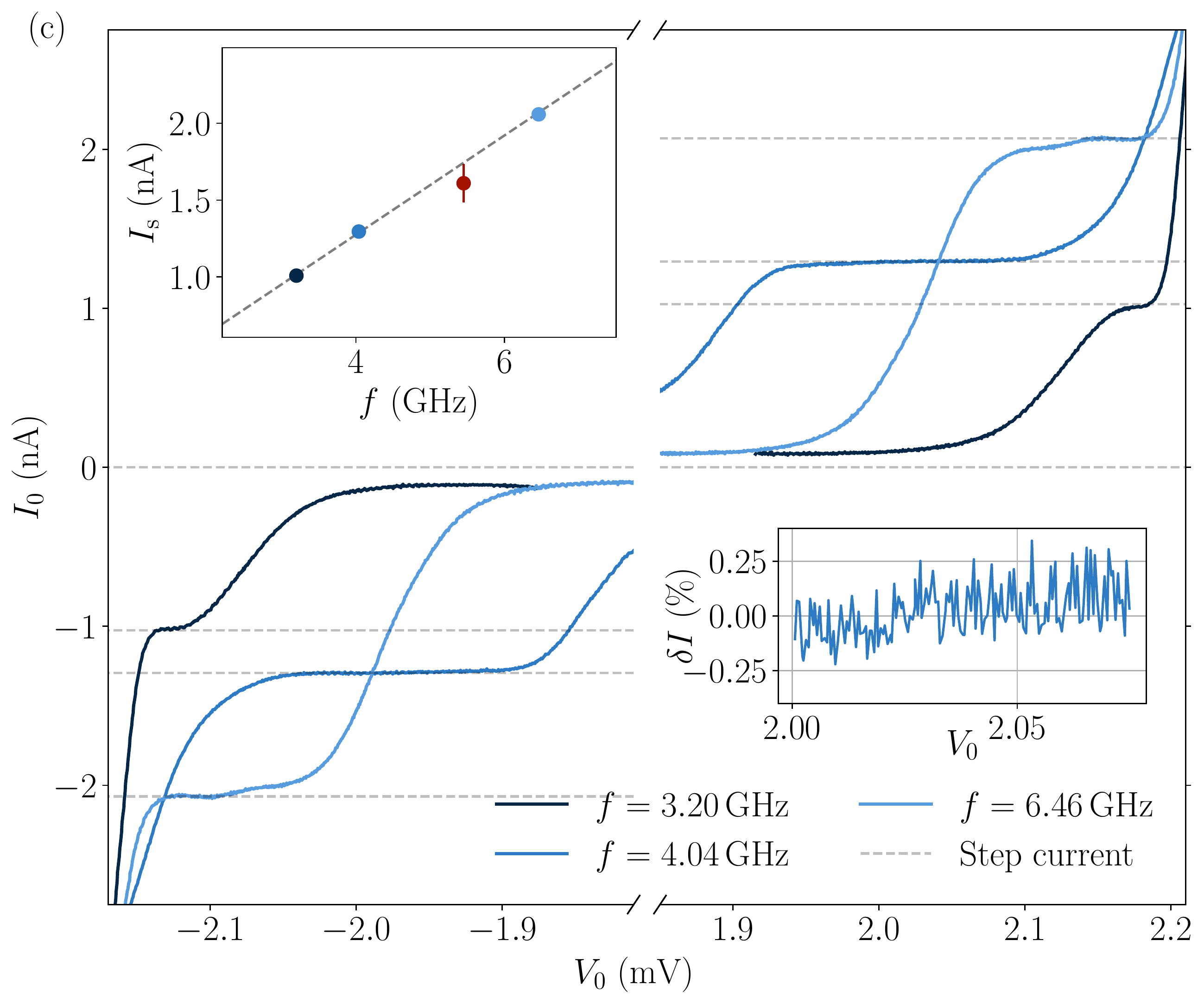}
	\caption{Current plateaux at different microwave tone frequency and power. (a) Average plateaux current at different microwave powers $p$ for three separate voltages. (b) Lock-in measurement collected by modulating the microwave power $p$ and detecting the corresponding current variation at $V_0=2.4\,\mathrm{mV}$. (c) For three different values of $f$ we follow the procedure described in the text to observe current plateaux whose amplitude correspond to the one of the first dual Shapiro step, which is shown as a grey dashed line. The inset shows the relation between the step current $I_s$ and tone frequency, and the fitted slope is consisted with $2e$. The red point's IV curve was omitted for clarity (see the main text and the Methods for further details). }
	\label{fig:steps}
	\end{figure*}

We now study the behaviour of the Bloch array in the presence of an additional applied microwave tone. In Fig.\,\ref{fig:pfsweep}a we display the microwave transmission of the Bloch array near the  mode doublet at 4.1\,GHz for different frequencies, $f$, of the microwave tone with power $p=-5.0\,\mathrm{dBm}$.  We observe that the array odd mode is influenced by the microwave tone, and drifts away from $f$, in a mode repulsion fashion. A mode synchronous with the pump appears when the frequency exceeds $\simeq4.02\,\mathrm{GHz}$, and fades around $f\simeq4.11\,\mathrm{GHz}$. 
This behaviour is further studied by varying $p$ at $f=4.04\,\mathrm{GHz}$, as reported in Fig.\,\ref{fig:pfsweep}b. At low power the array is essentially unaltered by the tone, until $p$ reaches -20\,dBm and the spectrum shows a profile corresponding to a Fano resonance, whose maximum-to-minimum separation increases with $p$. 
The most noteworthy feature is at $p\simeq-5\,\mathrm{dBm}$, which is presented in Fig.\,\ref{fig:pfsweep}c.
We clearly observe a microwave mode with a Lorentzian shape and linewidth of about $1\,\mathrm{MHz}$, centered at $f$, which disappears at higher $p$. 
%This observation is consistent with the onset of a Bloch mode synchronised with the external microwave tone\cite{lenz_bloch_1999}.
We refer to this feature as a Bloch mode, as it is consistent with Bloch oscillations synchronised with the external tone\cite{lenz_bloch_1999}.

Focusing on this microwave synchronised Bloch mode ($f=4.04\,\mathrm{GHz}$, $p=-5.0\,\mathrm{dBm}$), we now examine the IV characteristic of the Bloch array under microwave irradiation as displayed in Fig.\,\ref{fig:pfsweep}d.  The IV presents a wide central region with no current or supercurrent peaks. Outside this central part, we observe that for both positive and negative voltage, current plateaux form before the $2\Delta$ current peaks. The current corresponding to their flat part is $2ef$ within a 0.1\% accuracy. 
These observations are supported by simulations (see Methods) showing that the steps should emerge as a flat current region right before the upward slope of the $2\Delta$ peaks.
%\comment{NC-should we mention that the high power is the responsible for the absence of the first three current peaks?}

Figure \ref{fig:steps} displays a more detailed investigation of these current steps. 
In Fig.\,\ref{fig:steps}a we display the heights of three different current plateaux versus $p$, which have a saturation point at $p=-5\,\mathrm{dBm}$, and $I_0=2ef$.
The tone power $p$ influences the height of the step and quantized flat steps are obtained only for a given power range, in a fashion very similar to what is observed for standard Shapiro steps \cite{Calvez.2019}. To further assess the stability of the step, we modulate the microwave power $p$ and measure the corresponding $I_0$ variation with a lock-in amplifier. The result is displayed in Fig.\,\ref{fig:steps}b and shows that $dI_0/dp$ falls to the noise floor at $p\simeq-5\,\mathrm{dBm}$, suggesting a current that is independent of the microwave power.
	
Microwave power is delivered to the UJJ when the tone is on resonance with the odd microwave mode of the array. Utilising microwave spectroscopy, we identify other suitable working points (see Methods), corresponding to different mode doublets of the Bloch array. In Fig.\,\ref{fig:steps}c, we display the lowest bias voltage current step common to three additional frequencies. The parameters and additional measurements used to obtain these curves are detailed in the Methods section.
For a total of four mode frequencies, we observe that $I_0$ flattens at the corresponding $2ef$ for $f=3.20\,\mathrm{GHz},\,4.04\,\mathrm{GHz},\,5.45\,\mathrm{GHz}$ and $6.46\,\mathrm{GHz}$. 
Finally, the inset of Fig.\,\ref{fig:steps}c shows the frequency dependence of $I_s$, the average current on the flat part of each step. Through the slope of the fitting line we can estimate a Cooper pair's charge of $(3.23\pm0.07)\times10^{-19}\,\mathrm{C}$. 
Even though the measurement was not optimised, the average deviation from the expected current $\delta I =I_0/2ef - 1$ on the 4\,GHz step is $\langle \delta I^2\rangle^{1/2} \le0.001$, limited by the setup noise in an averaging time of one minute (see inset of Fig.\,\ref{fig:steps}c).

The environment of the UJJ, formed by the Josephson junction arrays, does more than provide the required high impedance for charge localisation. As we have discussed, the switching of the junctions in the array absorbs the majority of the applied voltage, allowing the UJJ to remain in the superconducting state even at high $V_0$. Furthermore, the microwave response of the device allows us to selectively apply a microwave drive to the UJJ, causing the formation of the Bloch mode and the corresponding onset of current steps. In contrast, no effect is observed when driving the even mode which does not couple to the UJJ. 
% the even modes are not coupled to the UJJ, and subsequently no Bloch oscillations are observed, and no steps are formed. In contrast, for nearly every odd mode we observe the formation of current steps, further supporting the UJJ's critical role in the observed physics.   %This is further supported by the fact that we only observe the formation of steps when the odd mode of the microwave doublet is driven with a tone  preventing

Our work highlights how the Bloch array can be harnessed to engineer dissipation, thereby resolving charge at the expense of phase fluctuations, a regime dual of what is commonly achieved in Josephson circuits (localised phase and large charge fluctuations). We take advantage of this new configuration and observe quantised current steps: dual Shapiro steps.
The quantum metrology electrical triangle links current, voltage, and frequency via three quantum effects. Voltage and frequency can be related via  the AC Josephson effect, while current and voltage via  the quantum Hall effect. Here we study the third edge of the triangle by relating current and frequency via the charge $2e$ of a Cooper pair. This first step in this direction indicates the rich physics of the Bloch array will have to be fully mastered in the process of developing the dual Josephson effect into a metrological definition of the Ampère\cite{newsi,pekola_single-electron_2013}.
This also opens exciting opportunities from the perspective of fundamental physics via tests of quantum electrodynamics through the closure of the triangle\cite{sailer_measurement_2022} or further exploration of phase charge duality in Josephson devices.

%\section{Acknowledgements}
\textit{Acknowledgements} -- The help of Denis Basko is deeply acknowledged and appreciated.
The support of the superconducting circuits team of Institut Néel is warmly acknowledged.
The authors are also grateful to Joe Aumentado, Michel Devoret, Tim Duty, Serge Florens, David Haviland, Julien Renard, Benjamin Sacepe and Izak Snyman for the fruitful discussion and comments on our work. 
Furthermore, our gratitude goes to the members of the Triangle consortium, namely Philippe Joyez, Çağlar Girit, Cristiano Ciuti, Helene Le Sueur and Alexander Wagner, for the valuable discussion and insights. 
The samples were fabricated in the clean room facility of Institute Néel, Grenoble; we would like to thank all the staff for help with the devices fabrication. 
We would like to acknowledge Eric Eyraud for his help in the installation of the experimental setup. 
This work was supported by the French National Research Agency (ANR) in the framework of the TRIANGLE project (ANR-20-CE47-0011). N.C. is supported by the European Union’s Horizon 2020 research and innovation program under the Marie Skłodowska-Curie grant agreement QMET No. 101029189. K.W.M acknowledges support from NSF Grant No. PHY-1752844 (CAREER), AFOSR MURI Grant No. FA9550-21-1-0202, and ONR Grant No. N00014- 21-1-2630.

During the writing of this work, we became aware that the group of Evgeni Il'ichev and Oleg Astafiev observed quantised current steps in a superconducting nanowires system. We would like to thank them for the discussion and their comments.

\bibliographystyle{unsrt}
\bibliography{dShap}

\begin{thebibliography}{10}
\bibitem{largesmall}
The array's first junctions reported in the figure are slightly larger than the
  ones in the rest of the device (not shown). This is not expected to result in
  any difference.
\bibitem{QMET_repo}
The DOI for a data repository will be inserted upon the acceptance of the
  manuscript.
\end{thebibliography}

%==============================================================================%
\newpage
\
\newpage

\section{Methods}
\label{sec:methods}
The Methods sections are organised as follows. In the first two sections we describe the phase charge duality and a semiclassial model that can give intuition into the observations, and simulation results from a minimal a RCSJ model.
In the third section we detail the device fabrication, parameters and experimental measurement setup. 
The fourth section presents the full IV curve of the device as well as its flux dependence, while the fifth one describes the microwave properties of the device. The last two sections display the characterisation of the microwave modes used for the observation of the current steps and their power and flux dependence.

\subsection{Semiclassical model and exact duality}
\label{sec:methods_theory}
\begin{figure}[b]
    \centering
    \includegraphics[width=.5\textwidth]{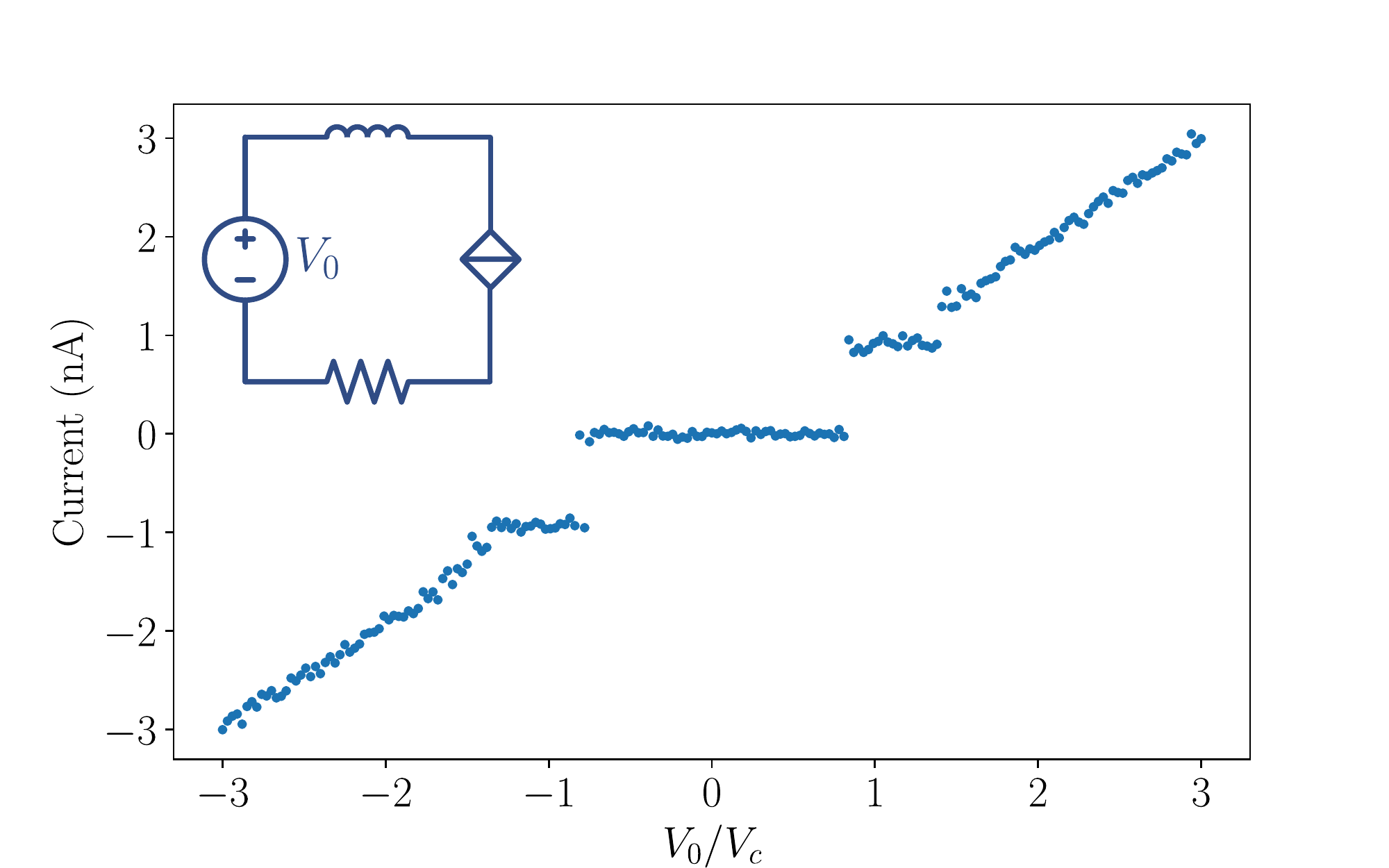}
    \caption{Simulation of the CJLR circuit under microwave irradiation with experimental noise. This IV curves are used to determine how clear are the steps for different parameters combinations and do not resemble the actual steps observed in our device. The inset is the simulated circuit, where the diamond is the non-linear capacitor ($\cos(\frac{\pi}{e}Q)$ element).}
    \label{fig:duals_sim}
\end{figure}
The two energy scales in a Josephson junction are the Josephson energy $E_J$ and the charging energy $E_C$. A large $E_J/E_C$ ratio is used, for example for the Josephson voltage standard \cite{doi:10.1063/1.1289507} or in transmon qubits \cite{PhysRevA.76.042319,PhysRevB.77.180502}, where an almost classical behaviour of the superconducting phase is favourable. When $E_J\lesssim E_C$, the charge of a single electron is relevant, and quantum phase fluctuations are not negligible. In this regime, charge dynamics can be dual to the Josephson phase dynamics, provided that the environmental impedance and inductance $L$ are large\cite{guichard_phase-charge_2010,di_marco_quantum_2015,arndt_dual_2018}.
Eq.\,(\ref{eq:hamiltonian}) describes a particle in a one-dimensional lattice, whose spectrum consists of Bloch bands\cite{Ashcroft:102652}, as shown in Fig.\,\ref{fig:triangle}. If the dynamics is restricted to the first Bloch band, and the bath response is ohmic with resistance $R$% and corresponds to that of a resistor $R$
\cite{guichard_phase-charge_2010}, the equation of motion for the charge is given by
    \begin{equation}
    L\ddot{Q} + R\dot{Q} + V_c\sin\left( \frac{\pi}{e} Q\right) = V_0+V_1,
    \label{eq:motion}
\end{equation}
where $V_c$ is the critical voltage of the junction, $V_0+V_1$ are the DC and AC applied voltages, respectively, and the shape of the first Bloch band is approximated as $\cos(\pi Q/e)$. Eq.\,\ref{eq:motion} is the exact dual to the equation of motion of a large $E_J/E_C$ ratio JJ.
While Eq.\,\ref{eq:motion} does not accurately model the complex physics of the Bloch array, it does provide useful intuition and serves as a guide for choosing the system parameters.

Following the work of Guichard and Hekking \cite{guichard_phase-charge_2010}, we calculate the time evolution of the system for a fixed voltage, allowing us to produce the IV characteristic of the circuit as reported in Fig.\,\ref{fig:duals_sim}.
The CJRL system under consideration is reported in the lumped element schematic in the inset of Fig.\,\ref{fig:duals_sim}. The usual phase dynamics associated with the Josephson tunneling is replaced by its dual counterpart: the quasicharge dynamics. 
In this way, instead of the voltage Shapiro steps, one can expect quantised current steps. As with prior work\cite{guichard_phase-charge_2010}, here the signal is perturbed with the voltage noise of the resistor and of the readout amplifier.
The circuit parameters are the same as the ones in the main text. 
%The figure clearly displays a current plateau starting at $V_0\simeq V_c$. 

\subsection{RCSJ toy model}
\label{sec:rcsj}
\begin{figure*}
    \centering
    \includegraphics[width=\textwidth]{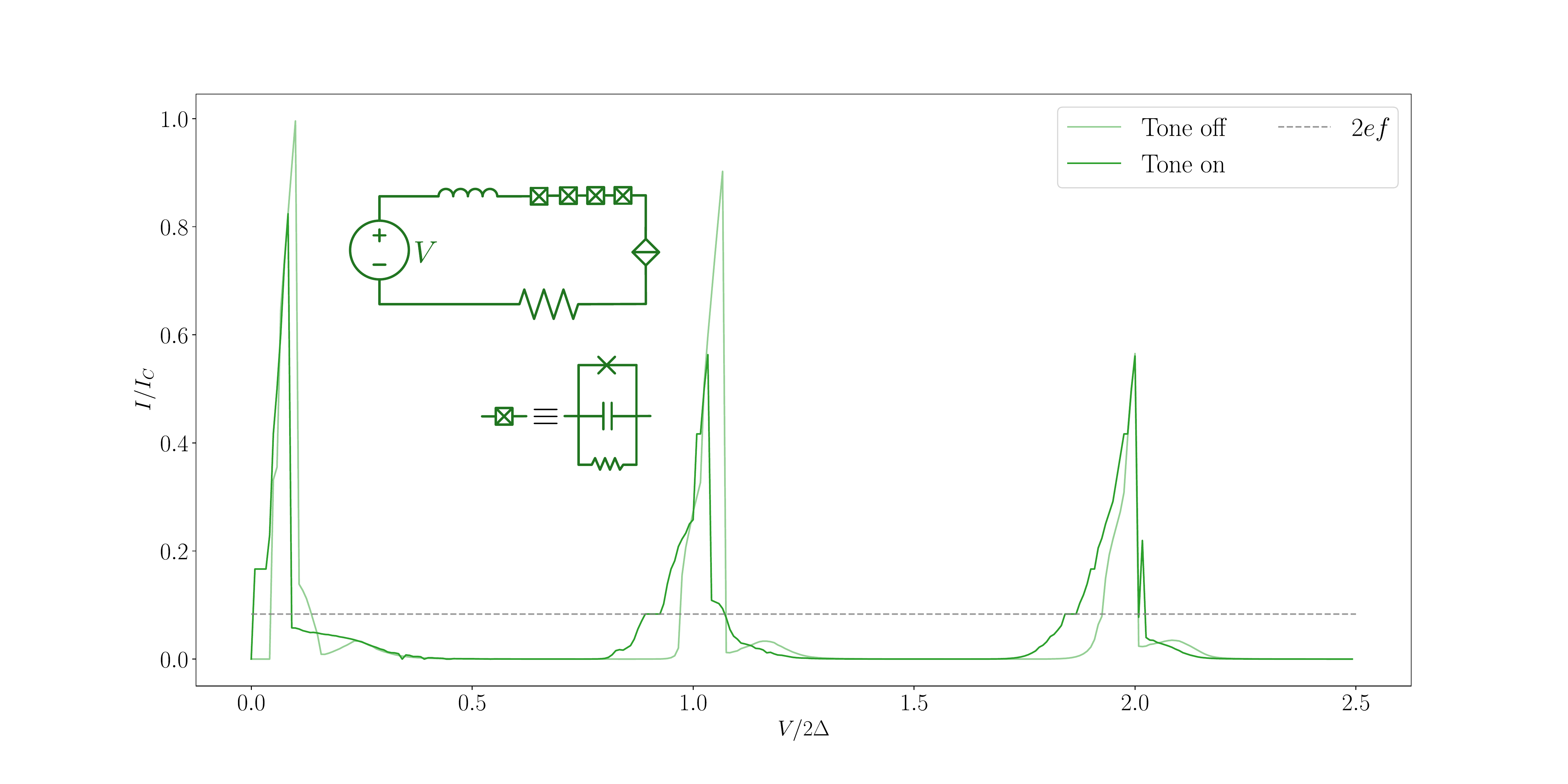}
    \caption{Simulated IV curve of a chain of four junctions in series with an inductance, a resistor, and a non-linear capacitor with and without an AC drive of frequency $f$. Current plateaux of height $2ef$ are observed before each $2\Delta$ current peak in qualitative agreement with the experimental observations.}
    \label{fig:rcsj_sim}
\end{figure*}
The IV curve of the array is simulated by numerically solving the equations of motion of a chain of $N$ junctions in series with a non-linear capacitor, an inductance and a damping resistor under a voltage drive. Each junction is modeled using the resistively and capacitively shunted junction model (RCSJ). The total current $I_\mathrm{tot}$ through a junction is thus the sum of a capacitive current $I_C$, a resistive current $I_R$ and a Josephson current $I_J$:
%\begin{align}
%I_{tot} &= I_C+I_R+I_J \nonumber\\
 %&= C\ddot{\phi}_i + s\left(\dot{\phi}_i\right)\frac{1}{R_N}\dot{\phi}_i + \left(1-s\left(\dot{\phi}_i\right) \right)I_c \sin\left ( \frac{1}{\phi_0} \phi_i \right) \label{eq:rcsj_jj}
%\end{align}
\begin{align}
\begin{split}
    &I_\mathrm{tot} = I_C+I_R+I_J,  \label{eq:rcsj_jj}\\
    &I_C = C \ddot{\phi}_i, \\
    &I_R =s(\dot{\phi}_i)\frac{1}{R_N}\dot{\phi}_i,  \\
    &I_J = \left(1-s(\dot{\phi}_i) \right)I_c \sin\left ( \frac{\phi_i}{\phi_0} \right). 
\end{split}
\end{align}
In Eqs.\,(\ref{eq:rcsj_jj}), $\phi_i$ is the superconducting flux drop across the $i^\mathrm{th}$ junction, $R_N$ is its normal state resistance, $C$ is its capacitance and $I_c$ its critical current. We model the gap transition of the junction through the factor $s(\dot{\phi}_i)$ which goes smoothly from 0 when $|\dot{\phi}_i| < 2\Delta$ to 1 when $|\dot{\phi}_i| > 2\Delta$. We found that the exact shape of the function $s$ does not impact the resulting IV characteristics as long as the transition is sharp. All the results presented used a scaled sigmoid for the transition function.
%\begin{equation*}
%    s\left(\dot{\phi}_i\right) = \frac{1}{1+\exp\left(-10\left(e|\dot{\phi}_i| - 2\Delta\right)\right)}.
%\end{equation*}
The charge and its associated current $I_\mathrm{tot}=\dot{Q}$ are related to the bias voltage $V$ by an equation similar to (\ref{eq:motion})
\begin{equation}
    V =L\ddot{Q} + R \dot{Q} + V_C\sin\left( \frac{\pi}{e} Q\right) + \sum_i \dot{\phi}_i.
    \label{eq:rcsj_voltagelaw}
\end{equation}
We then solve Eq.\,(\ref{eq:rcsj_jj}) and (\ref{eq:rcsj_voltagelaw}) using a Runge-Kutta method of order four with a time-step $dt=0.02\,RC$. We then average the current $\langle I_\mathrm{tot}\rangle$ to obtain its DC component and plot the IV characteristic.  

Due to the complexity of simulating the response of $N_\mathrm{a} = 3500$ JJs, we focus on a system with $N=4$ junctions embedded in a circuit with inductance $L=5\,\mathrm{\mu H}$ which captures the inductance of the remaining JJs that form the superinductance. 

Fig.\,\ref{fig:rcsj_sim} displays the IV characteristic for the chosen parameters, with and without an applied microwave tone. Many features that we observe in the experimental IV curve are qualitatively reproduced by the simulation. First, we observe current peaks, spaced by $2\Delta$, corresponding to the successive switching of junctions in the array. Second, when the microwave tone is applied we observe the formation of current plateaux just before the voltages at which the current peaks occur. These current plateaux are repeated for each current peak. 

Other features of this simple simulation do not qualitatively match the experimental IV curves. 
% peaks height
Whereas the height of the simulated current peaks is of order $I_c$, and monotonically decreasing with voltage, the experimental current peaks are all well below the estimated $I_c$, with heights suppressed for small $V_0$. This qualitative difference points to aspects of the UJJ, which is simplified in the model.
% plateau width
The width of the simulated plateaux appear to be of order $V_c$, while experimentally we measure different widths, as is shown in Fig.\,\ref{fig:steps}. 
%Being the one obtained with the lower microwave power (see Table \ref{tab:fp}), the $3.20\,\mathrm{GHz}$ step matches the expected width. This indicates that in this condition the tone does not interfere with other features of the array (i.\,e. degrading the superconductivity, or switching some junctions), and the observation of steps with a weaker tone is preferable.

\subsection{Device and experimental setup}
\label{sec:methods_dev}
The device consists of Al/AlO$_x$/Al junctions in a microstripline geometry, fabricated with the shadow evaporation technique and electron beam lithography on a fused silica wafer. The substrate choice is related to the minimisation of ground capacitances which, being proportional to the dielectric constant, are reduced of about four-folds with respect to silicon. Reduced ground capacitance and large inductance produce a high impedance and prevent Landau-Zener transitions between Bloch bands which might suppress the amplitude of Bloch oscillations \cite{arndt_dual_2018}. Details on the fabrication procedure can be found in the data repository of this work\cite{QMET_repo}.

The superinductance is formed from JJs with capacitance  $C_\mathrm{a}\simeq45\,\mathrm{fF}$ and critical current $I_c\simeq 150\,\mathrm{nA}$, estimated from their area by using the expected capacitance per unit of area $\sigma_C=45\,\mathrm{fF/\mu m^2}$ and critical current density of $\sigma_I = 15\,\mathrm{A/cm^2}$, respectively. 
Each UJJ has an area of $0.04\,\mathrm{\mu m^2}$, from which we can similarly estimate a capacitance of $C=1.8\,\mathrm{fF}$ and a critical current of $6\,\mathrm{nA}$.

\begin{figure}
    \centering
    \includegraphics[width=.5\textwidth]{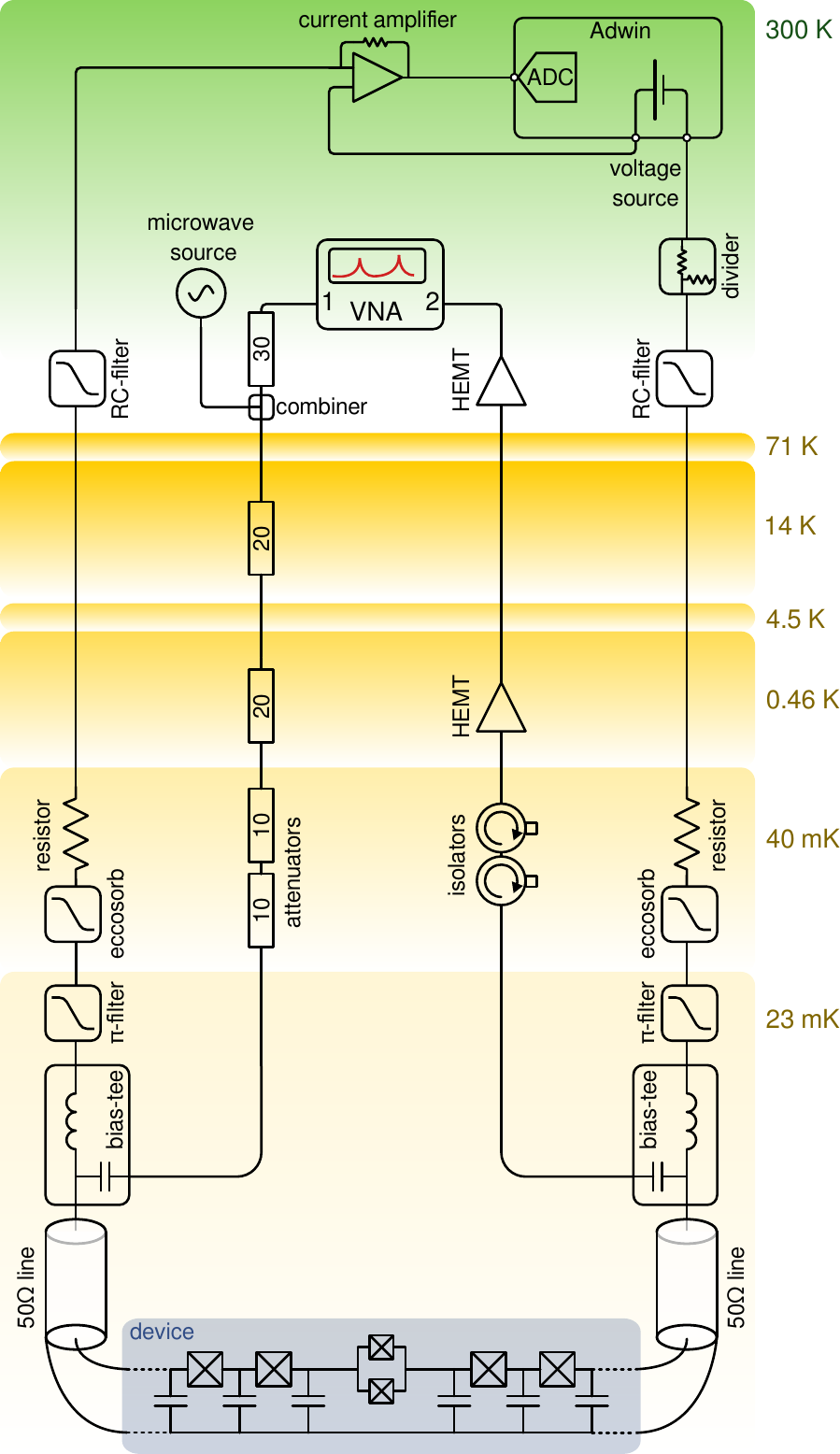}
    \caption{Experimental setup used in this work. The yellow boxes show the different stages of the dilution refrigerator with their temperatures on the right side, while the part of the apparatus at room temperature is in green. The device is in gray at the bottom of the figure, the reader is referred to the main text for its description. }
    \label{fig:setup}
\end{figure}

A detailed schematic of the setup used in this work is reported in Fig.\,\ref{fig:setup}.
The experimental apparatus can simultaneously perform DC and RF measurements, as is shown in Fig.\,\ref{fig:circs}c. This is achieved using two off-chip bias-tees that decouple kHz frequency measurements from the ones in the GHz range. The two ports of the Bloch array are connected to the bias-tees via $50\,\mathrm{\Omega}$-matched lines. At the lowest temperature stage, the DC lines feature $45\,\mathrm{MHz}$ $\pi$-filters, $50\,\Omega$ damping resistors in series and Eccosorb filters. Further $RC$ and $\pi$-filters are applied at room temperature, which are the main source of damping in the circuit. The current $I_0$ is amplified with a low noise transimpedance amplifier, while the readout and voltage bias are performed with a  real-time analog-to-digital converter (Adwin-Gold II).

\begin{figure*}
	\centering
	\includegraphics[width=.49\textwidth]{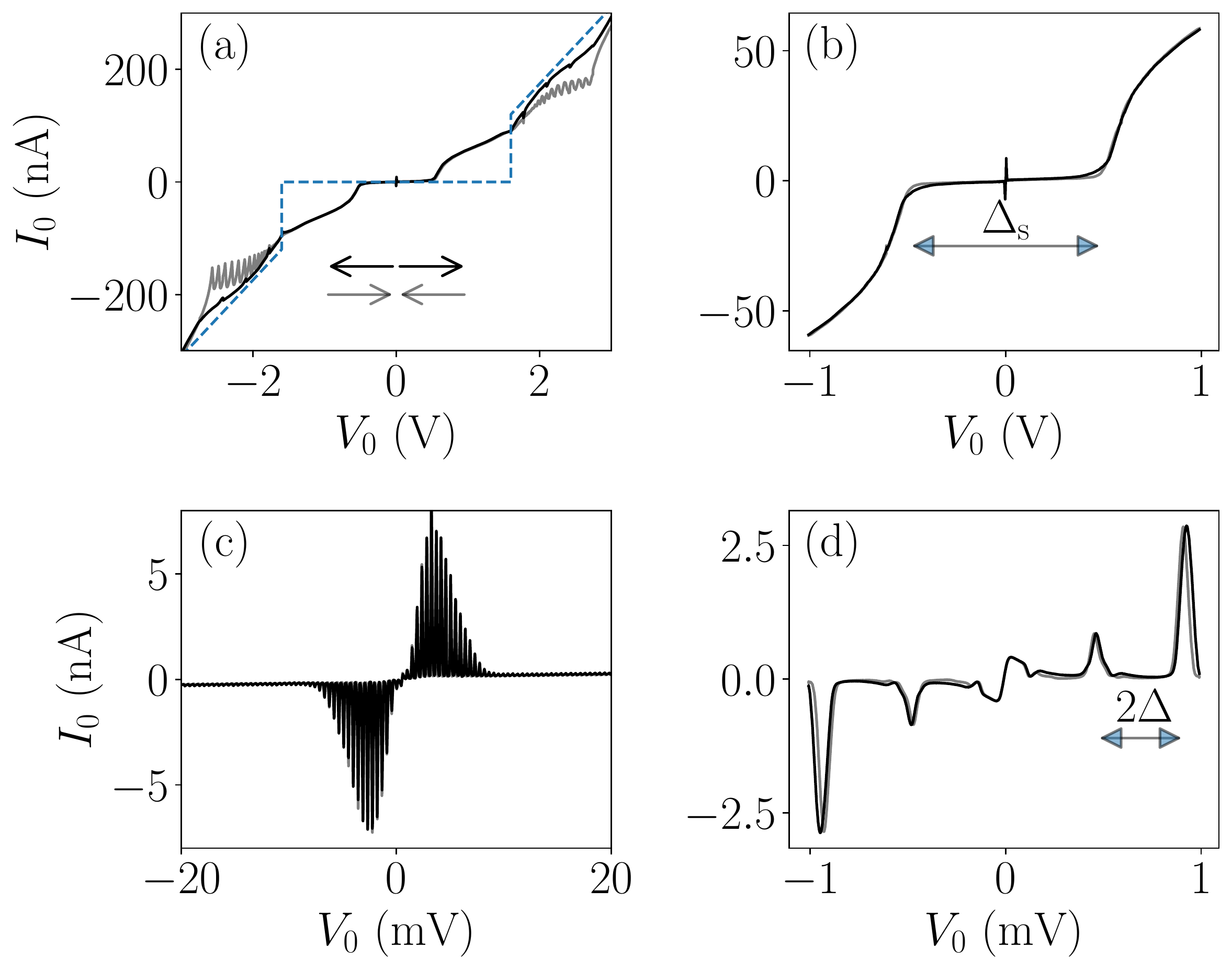}   
	\includegraphics[width=.49\textwidth]{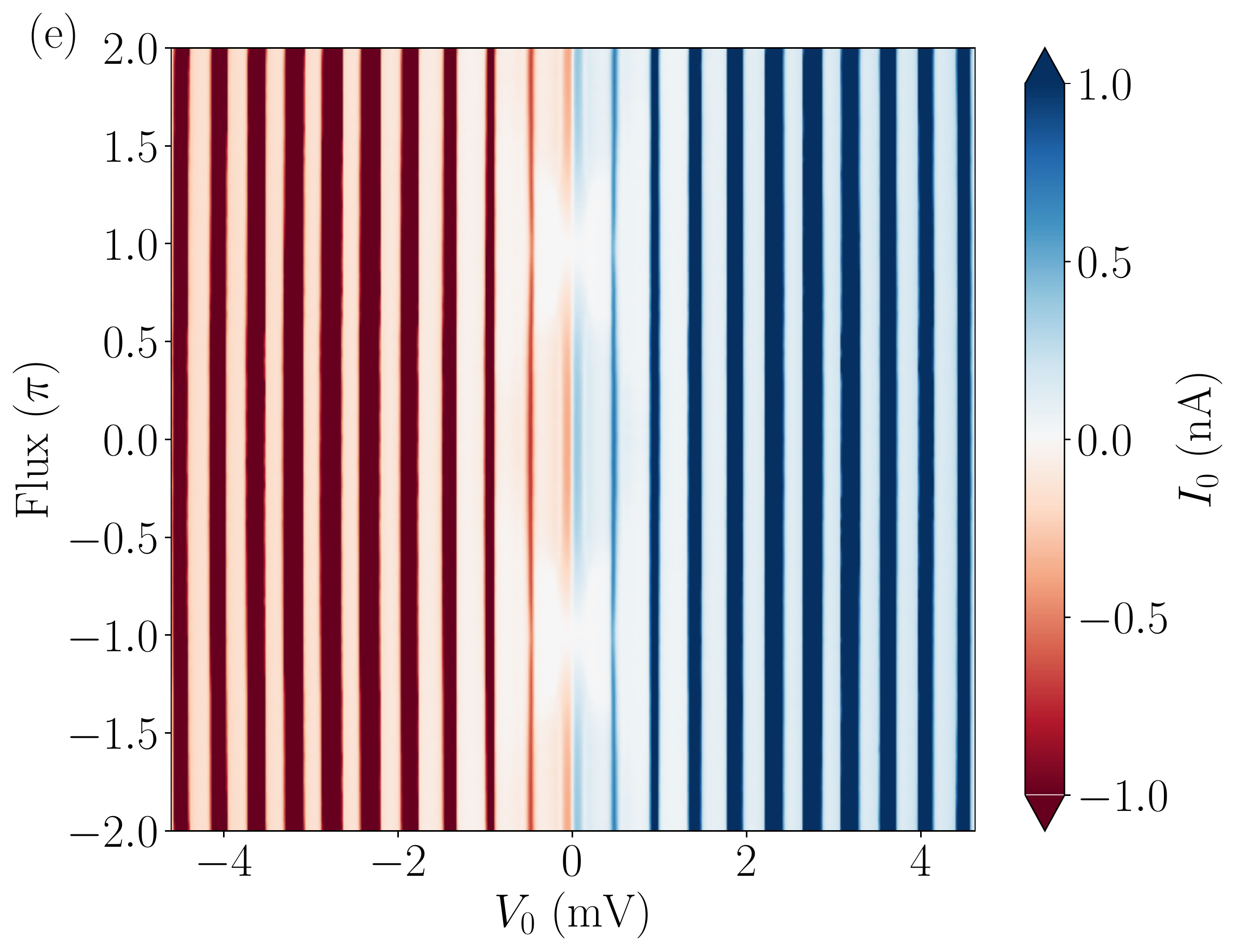}
	\caption{Full IV curve of the Bloch array. Plots (a) to (d) show the IV curve at different scales. The black (grey) line indicates increasing (decreasing) $|V_0|$, as shown by the arrows in (a), while in the same plot the blue dashed line is a guide for the eye. The voltage scale is gradually reduced from (a) to (d); discrepancies between the current amplitude in the plots are related to measurement conditions. Finally, (e) shows the flux dependence of the Bloch array's IV characteristic at low voltages.}
	\label{fig:fulliv}
\end{figure*}

The IV curves are collected by biasing the device with triangular voltage ramps with period ranging from 10 ms to 10 s, depending on the measurement conditions and voltage range.
To apply a magnetic field on the sample we employ a supeconducting coil anchored to the sample holder. 
The measured periodicity in the IV response agrees with what is estimated considering the SQUID area and the magnetic field produced by the coil. 
Microwaves are applied to the sample via a a high frequency line with an attenuation of about 60\,dB, which is connected to one of the bias-tees, while spectroscopic measurements are performed with extra 30\,dB attenuation. The second bias-tee's high frequency port is connected to an isolator and then to a high electron mobility transistor (HEMT) amplifier working in the 4-8\,GHz frequency range and with a noise temperature of about $5\,\mathrm{K}$. Further amplification is performed at room temperature, before reading out the signal with a vector network analyser.  The setup is housed in an inverted dilution refrigerator with a base temperature of $23\,\mathrm{mK}$.

% Full IV curve and flux dependence
\subsection{Full IV curve and flux dependence}
\label{sec:fulliv}
The current-voltage characteristics of the Bloch array are reported in Fig.\,\ref{fig:fulliv}.
Figure \ref{fig:fulliv}a shows the IV curves over the largest voltage span, with switching to the resistive branch of the whole array at a voltage of about $2N_\mathrm{a}\Delta\simeq\pm1.6\,\mathrm{V}$, marked by a sharp peak in the device conductance. From this, we estimate $\Delta=0.23\,\mathrm{mV}$, compatible with the superconducting gap of aluminium thin films \cite{Court_2007}.
For applied voltages $V_0\sim 4$\,V, we estimate the normal state resistance of each JJ in the array $R_\mathrm{a}=7.75\,\mathrm{M\Omega}/N_\mathrm{a}=2.21\,\mathrm{k\Omega}$. Since, the resistance of the UJJ is expected to be $\sim 20\,\mathrm{k}\Omega$, its contribution to the total array resistance is negligible. % is negligible since it contributes as one part in $3500=N_\mathrm{a}$.
Using Ambegaokar–Baratoff formula, $I_c = \pi\Delta / 2R_\mathrm{a}$,  we estimate the critical current of the large junctions $I_c\simeq164\,\mathrm{nA}$, compatible with the estimation obtained with SEM imaging.
%\begin{equation}
 %   I_c = \pi\Delta / 2R_\mathrm{a}
 %   \label{eq:ambegaokar}
%\end{equation}
%   resulting in $I_c\simeq164\,\mathrm{nA}$, which is compatible with the estimations obtained with SEM imaging. %\comment{$I_c$ is used multiple times throughout the paper, shall we give it a different name every time? It is calculated in different ways but it's always the same quantity.}
%

At applied voltages $V_0<\pm1.6$\,V the IV has a more resistive branch and a central plateau of about $\Delta_\mathrm{s}\simeq 1\,\mathrm{V}$, where its resistance is larger than $100\,\mathrm{M\Omega}$, as shown in Fig.\,\ref{fig:fulliv}b. This is in agreement with prior measurements and theory \cite{ergul_phase_2013}.
% full IV
At a lower voltage scale, see Figs.\,\ref{fig:dcrf}a and \ref{fig:fulliv}c, more features appear \cite{haviland_superconducting_2000,dolata_single-charge_2005}, whose explanation is reported in the main text. Current peaks are present when the voltage is an integer multiple of $2\Delta$, and their spacing is understood in terms of a sub-gap to resistive switching mechanism as explained in the main text.

%However, the supercurrent associated with the linear junctions should be of order $I_c\sim150\,\mathrm{nA}$, while we measure a maximum peak current of about $10\,\mathrm{nA}$ at $V_0 \simeq 7(2\Delta)$.
%However, both previous measurements \cite{haviland_superconducting_2000,dolata_single-charge_2005,weissl:tel-01369020} and our simulations show how the maximum current of the peaks should monotonically decrease for increasing $V_0$.
%Instead, the peak current decreases before and after this value creating a regular envelope; which we attribute to the UJJ influencing the neighboring junctions.% causing them to switch first. This is consistent with microwave measurements as discuss in the following section, in particular in Fig.\,\ref{fig:tt_abcd}c.

Figure \ref{fig:fulliv}d displays the IV curve for the finest voltage scale. For $V_0<\Delta$ we detect a low-voltage-current with non-trivial shape which we associate with a combination of supercurrent and inelastic Cooper pair tunnelling \cite{ingold_charge_1992,PhysRevLett.115.027004,PhysRevLett.106.217005}.
Small applied magnetic fields are expected to tune the $E_J/E_C$ ratio of the UJJ, while the superinductances, being formed of single junctions are to first order not affected.
Figure\,\ref{fig:fulliv}e shows the flux dependence of the low voltage IV curve of the Bloch array, where at $\pm\pi$ flux we observe suppression of the low-voltage-current\cite{PhysRevLett.115.027004,PhysRevLett.106.217005}. 

\subsection{Microwave measurements of the Bloch array}
% Two-tone spectroscopy and ABCD
%\begin{figure*}
%    \centering
%    \includegraphics[width=.75\textwidth]{img/twotones.pdf}   
%    \includegraphics[width=.75\textwidth]{img/s21_abcd_b.pdf}
%    \caption{Microwave characterisation of the Bloch array. (a) Two-tones measurement with readout on the 4.13\,GHz mode analysed in the main text, showing the different array modes and the plasma frequency of about 16\,GHz. (b) \textit{ABCD} matrix simulation of the Bloch array with and without UJJ. Without UJJ the circuit behaves as 3500 junctions-long JJ transmission line. The addition of a UJJ, schematised as a capacitance at the center of the array, yields the measured microwave transmission.}
%    \label{fig:tt_abcd}
%\end{figure*}

Microwave spectroscopy is used to measure the resonances of the Bloch array \cite{puertas_martinez_tunable_2019}. In particular, we identify several peak doublets, which can be further studied with two-tone measurements \cite{Krupko.2018}. We probe the transmission of the array near one of the observed resonances at 4.13\,GHz and sweep the power and frequency of an additional applied tone. When the tone is resonant with an additional mode of the array the transmission at the probe mode is affected. From these data we observe a free spectral range of about $1\,\mathrm{GHz}$ between two peaks doublets, and a JJ plasma frequency of 16\,GHz, beyond which no modes are observed. From these measurements, we obtain the JJ inductance $L_\mathrm{a}\simeq 1.9\,\mathrm{nH}$, capacitance $C_\mathrm{a}\simeq45\,\mathrm{fF}$, and ground capacitance $C_\mathrm{g}\simeq7\times10^{-4}C_\mathrm{a}=0.3\,\mathrm{aF}$.
These results are in agreement with the ones obtained by measuring the dispersion relation of a JJ array resonator fabricated in the same batch.

We then obtain the two superinductance's impedance and inductance $\sqrt{L_\mathrm{a}/C_\mathrm{g}} = \mathrm{8.0\,k\Omega}$ and  $N_\mathrm{a}L_\mathrm{a}/2 = 3.3\,\mathrm{\mu H}$. The critical current of the large JJs, $I_c= \hbar/2e L_\mathrm{a} \simeq 173\,\mathrm{nA}$ is in good agreement with the values estimated in the previous section.
The spectroscopic data also give information on the UJJs. Referring to Fig.\,\ref{fig:dcrf}c, the observed modes splitting yields a capacitance of about $1\mathrm{fF}\simeq C/2$, in agreement with the value estimated through the ultrasmall junction's area.

\begin{figure*}
    \centering
    \includegraphics[width=\textwidth]{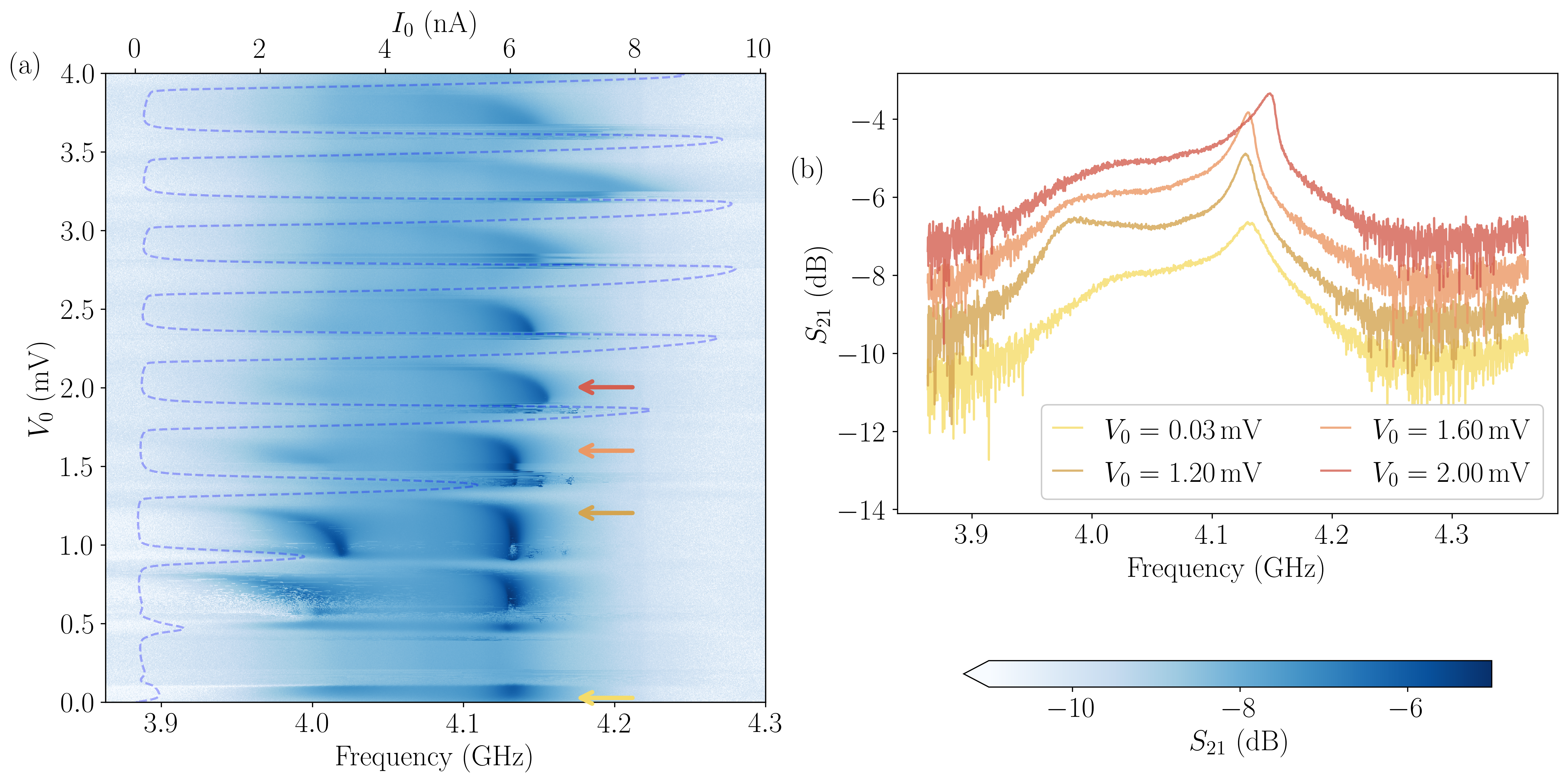}
    \caption{Plot (a) is the variation of the transmission with respect to $V_0$, where the IV was superimposed to the plot as a dashed line for illustrative purposes. In (b) we show slices of (a) at different voltages, where a 1\,dB offset for each curve is added for clarity.}
    \label{fig:s21_dc}
\end{figure*}

The effect of $V_0$ on the $S_{21}$ of this mode doublet is displayed in Fig.\,\ref{fig:s21_dc}. Here, we see how a current flowing in the Bloch array degrades the microwave response of both modes, reducing the transmission. The microwave mode doublet is recovered when $I_0$ is close to zero, further confirming the interpretation of the IV curve reported in the main text.

\begin{figure*}
    \centering
    \includegraphics[width=.95\textwidth]{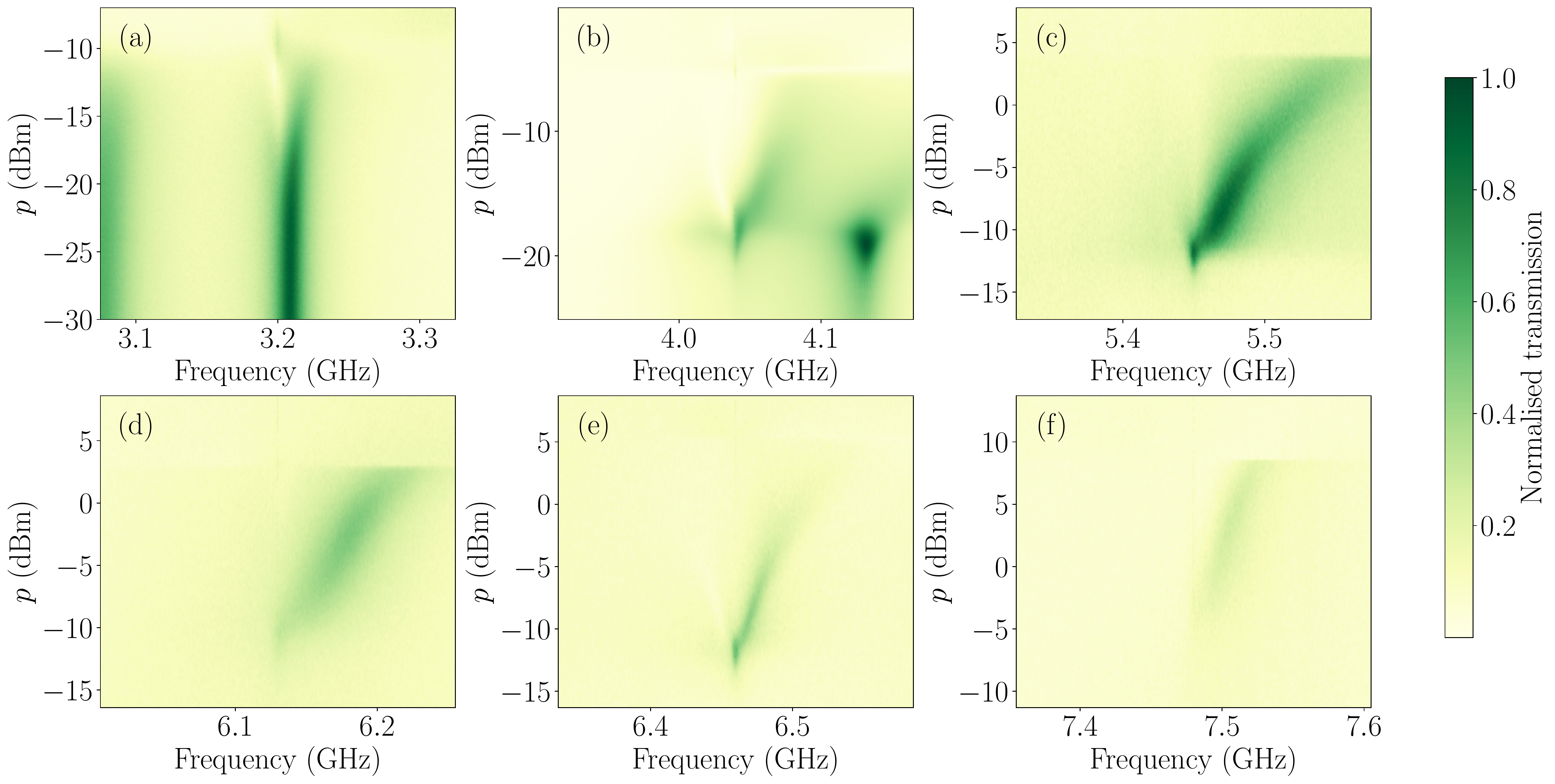}
    \caption{Power sweep on the peak doublet for some of the modes identified in the Bloch array, ranging from (a) at 3.2\,GHz to (f) at 7.5\,GHz. These measurements are used to identify the power to use to obtain the steps.}
    \label{fig:psweeps}
\end{figure*}

\subsection{Microwave modes used for observing current steps}
To observe the steps at different frequencies of the microwave tone, we identified the suitable peak doublets and their response to the microwave tone using the method displayed in Fig.\,\ref{fig:pfsweep}b. The corresponding power dependence for the other observed steps are displayed in Figure \ref{fig:psweeps}.  In each case, the characteristic shape of the power response guides the choice of $p$ to observe the steps. A summary of the identified frequencies and powers suitable for the emergence of the steps is reported in Table \,\ref{tab:fp}.
\begin{table}
	\begin{tabular}{ c | c  }
    $f$     	&	$p$	        \\
	\hline
	3.20\,GHz		&	-11.1\,dBm	\\
	4.04\,GHz		&	-5.0\,dBm	\\
	5.45\,GHz		&	2.8\,dBm	\\
	6.11\,GHz		&	Not identified	\\
	6.46\,GHz		&	3.8\,dBm	\\
	7.48\,GHz		&	Not identified	\\
	\hline
	\end{tabular}
	\caption{Microwave tone parameters used to observe the steps in Fig.\,\ref{fig:steps}, obtained through the measurements in Fig.\,\ref{fig:psweeps}.}
	\label{tab:fp}
\end{table}
Referring to Fig.\,\ref{fig:steps}c, we note that the 5\,GHz step is not displayed since the corresponding IV curve suffers from a much higher noise with respect to the other three. Repeated measurements show that with this tone frequency, the Bloch array experiences a significant $1/f$ noise which, shifting the position of the peaks during the measurement, compromises the averaging reliability. For this reason the averaging performed on this curve is extremely limited, leading to a higher current noise, as can be seen from the errorbar in the inset of Fig.\,\ref{fig:steps}c.

% \subsection{Power and flux dependence of the current steps}
% % Steps as a function of the tone power and flux
% \begin{figure*}
%     \centering    
%     \includegraphics[width=.47\textwidth]{img/steps_pd_2d_m.pdf}
%     \includegraphics[width=.49\textwidth]{img/ivs_flux_m.pdf}
%     \caption{(a) Power dependence of the first steps in the IV curve with $f=4.04\,\mathrm{GHz}$. With increasing $p$ the peaks closer to $V_0=0\,\mathrm{V}$ start to gradually broaden and fade, while the other remain almost unaffected. (b) Flux dependence of the current plateaux for the steps observed with $f=3.20\,\mathrm{GHz}$. }
%     \label{fig:ivs_flux}
% \end{figure*}
% In every measurement we observe how increasing $p$ gradually suppresses current peaks starting from the ones at lower voltage; as an example in Fig.\,\ref{fig:pfsweep}d several peaks are missing. 
% Figure \ref{fig:ivs_flux}a displays how the tone power affects the steps, which fade with the current peak when $p$ increases.

% Figure \ref{fig:ivs_flux}b displays the flux dependence of the current steps. Whereas the data of Fig.\,\ref{fig:pfsweep}d showing the steps are collected at zero flux, at small applied flux we observe a modulation of the IV characteristic, which affects the steps.  Steps are clearly viewed for small applied flux, with modulation corresponding to the flux quantum. At higher flux, the current peaks, and associated steps are significantly affected, likely due to the suppression of superconductivity and consequent variation of the optimal observation conditions.

\end{document}